# Approaching the Limits of Transparency and Conductivity in Graphitic Materials through Lithium Intercalation

Wenzhong Bao[(a), 1,2], Jiayu Wan[(a),2], Xiaogang Han[2], Xinghan Cai[1], Hongli Zhu[2], Dohun Kim,[1] Dakang Ma[3], Yunlu Xu,[3] Jeremy Munday[3], H. Dennis Drew[1], Michael S. Fuhrer[*, 1,4], Liangbing Hu[*, 2]

[(a)] These authors contributed equally to this work.
[1] Department of Physics, University of Maryland, College Park, MD 20742-4111, USA.
[2] Department of Materials Science and Engineering, University of Maryland, College Park, MD 20742-4111, USA.
[3] Department of Electrical and Computer Engineering, University of Maryland, College Park, MD 20742-4111, USA.
[4] School of Physics, Monash University, Victoria 3800, Australia.
*Corresponding Authors: binghu@umd.edu, michael.fuhrer@monash.edu

**Various bandstructure engineering methods have been studied to improve the performance of graphitic transparent conductors; however none demonstrated an increase of optical transmittance in the visible range. Here we measure *in situ* optical transmittance spectra and electrical transport properties of ultrathin-graphite (3-60 graphene layers) simultaneously via electrochemical lithiation/delithiation. Upon intercalation we observe an increase of both optical transmittance (up to twofold) and electrical conductivity (up to two orders of magnitude), strikingly different from other materials. Transmission as high as 91.7% with a sheet resistance of 3.0 Ω per square is achieved for 19-layer LiC$_6$, which corresponds to a figure of merit $\sigma_{dc}/\sigma_{opt}$ = 1400, significantly higher than any other continuous transparent electrodes. The unconventional modification of ultrathin-graphite optoelectronic properties is explained by the suppression of interband optical transitions and a small intraband Drude conductivity near the interband edge. Our techniques enable the investigation of other aspects of intercalation in nanostructures.**

**Introduction**

Two-dimensional (2D) graphene has attracted much interest in fundamental research and technological development due to its extraordinary electrical, mechanical, thermal and optical properties[1-6]. Recently, graphitic films (from monolayer graphene to ultrathin graphite) have been explored as candidates for flexible transparent electrodes for electronics and optoelectronics[7,8]. An excellent performance of 30 Ω per sq at 90% transmittance has been achieved using doped four-layer chemical-vapor-deposition (CVD) graphene.[9] Bulk materials with 2D layered structures such as graphite have also been studied and used extensively for electrochemical energy storage based on intercalation.[10-12] Fundamental studies on intercalation in graphite have been extensively carried out,[13] and nanostructured 2D materials have gained recent interest.[14] Reports on the intercalation of various species such as $FeCl_3$,[15,16] $Br$,[17] and $Ca$[18] in few-layer graphene (FLG) have offered a new route to designing and synthesizing graphene-based materials with novel conductive, magnetic, or superconductive properties.

It has long been known that the optical transmission of graphite increases upon metallization by intercalation with e.g. caesium[19]. This unusual property results from the unique band structure of the graphene layer; intercalation heavily dopes ultrathin graphite, shifting the Fermi level upward more than any other band engineering method[15,19-24], suppressing interband optical transitions due to Pauli blocking thus increasing transmittance of light in the visible range. The increase in optical transmittance is expected to be accompanied by an increase in conductivity since the carrier concentration increases upon intercalation, an ideal situation for conductive transparent films. All studies to date of doped graphene films as transparent electrodes, however, do not report increased transmission in the visible range. It also has been assumed by some researchers that the transmission of ultrathin graphite can never exceed that of undoped graphene of similar layer number[25]. Moreover, no studies of electrical conductivity and optical transmission have been carried out for lithium-intercalated ultrathin graphite.

Here we use *in situ* electronic and optical measurements to understand the electrochemical intercalation process and simultaneously measure the electrical conductivity and optical transmission of exfoliated ultrathin graphite crystallites ranging from 3-60 graphene layers in thickness. Upon intercalation we observe a large improvement in the optical transmittance, and at the same time a dramatic increase of sheet conductivity. The Dirac electronic bandstructure allows for very low electron-phonon resistivity even at relatively low carrier concentration[26], hence high DC conductivity is achieved with low optical conductivity below the visible range. In addition to elucidating the limits of conductivity and transparency in ultrathin graphite, we expect that the experimental techniques developed here will be broadly useful for studying the intercalation dynamics and correlated optoelectronic properties of other 2D nanomaterials that can be intercalated electrochemically.

## Results

**Devices for optoelectronic and transport measurements.** To simultaneously study the electrical and optical properties of Li-intercalated ultrathin graphite, we design a sandwich-structured cell with electrolyte (1 M $LiPF_6$ in w/w = 1/1 ethylene carbonate/diethyl carbonate) that is sealed by bottom and top layers of thin transparent glass. Ultrathin graphite and the lithium source are deposited on the bottom glass layer and connected to separate electrical contacts. Two types of devices are fabricated for our optoelectronic (Fig. 1a-c) and electrical transport (Fig. 1d-f) measurement. Details of the device fabrication are given in the Methods section. In this planar nanoscale half-cell battery (planar nano-battery), Li-metal is used as the counter electrode and ultrathin graphite as the working electrode. The intercalation process is controlled by a Bio-Logic SP-150 electrochemical workstation and the voltage of planar nano-battery can be measured simultaneously during the electrochemical Li-intercalation. The thickness of the ultrathin graphite is determined by an atomic force microscope (AFM) before cell capsulation.

***In situ* optoelectronic measurement.** The transmittance at a particular wavelength of pristine and Li-intercalated ultrathin graphite can be characterized by analyzing the grey-scale images acquired by transmission optical microscopy (Nikon Eclipse Ti-U) using a broadband light source (Thermo Oriel), a monochromator (Spex 500M), and a charge-coupled device (CCD) camera. A schematic of *in situ* transmittance measurement system is shown in Fig. 2a (also see Methods). A series of optical images (550 nm illumination) corresponding to different stages are shown in Fig. 2d-i, along with a schematic of the lattice structure of $LiC_6$ (Fig. 2h). A clear increase in the transmittance upon intercalation can be seen from Fig. 2e-g, as discussed below in detail. Furthermore, the optical transmittance change is highly reversible, as shown in Fig. 2i (also see Supplementary Fig. 3).

Our transparent planar nano-battery setup also allows further characterization using Raman microscopy. As shown in Fig. 2j we examined a series of *in situ* Raman spectra (Horiba Jobin Yvon with a 633 nm laser source), which correspond to different stages of Li-graphite intercalation, with more details discussed in the Supplementary Information. The result from our ultrathin graphite samples agrees well with previous studies of bulk samples[27], confirming Raman microscopy as one of the tools for differentiating lithiation stages in ultrathin graphite.

We also observed an *in situ* transmittance change of Li intercalated ultrathin graphite at different $Li_xC$ stages by charging the Li-graphite nano-batteries (Fig. 3a), with a constant charge current (Supplementary Information). The black line represents a typical voltage profile of the Li-graphite nano-battery and the red, green, blue open circles depict the transmittance (550 nm illumination) evolution of ultrathin graphite sheets (dotted regions in the inset of Fig. 3a) with different thicknesses during Li-graphite intercalation. The voltage initially drops rapidly with time until it reaches 0.8 V, where an obvious slope change in the voltage profile is observed. This is due to the decomposition of the electrolyte and a solid electrolyte interphase (SEI) formation[11,27]. No obvious change in the transmittance of the ultrathin graphite samples is observed for voltages greater than 0.2 V. A sudden increase in the transmittance (18-layer graphite from 74.4% to 77.2%, 38-layer graphite from 55.9% to 59.2%) occurs after 0.2 V, which we identify with the formation of $LiC_{36}$ (stage IV)[28] from ultrathin graphite sheets. From 0.2 V to 0.1 V, a gradual change in the transmittance is observed, presumably due to the formation of $LiC_{27}$ and $LiC_{18}$.[28] As time increases a second plateau appears in the voltage at ~0.1 V, which we identify with the formation of $LiC_{12}$ (stage II). At the end of the 0.1 V plateau we expect the entire sample has been converted to $LiC_{12}$, and the transmittance dramatically increased to 85.8% (18-layer) and 71.9% (38-layer). A third voltage plateau appears at a value of ~0.05 V, indicating the formation of $LiC_6$ (stage I). At this stage the transmittance of the 18-layer sample has increased to 90.9%, and the 38-layer sample has increased to 79.2%. Only two distinct stages are observed for the 3 layer sample (from 94.5% to 95.2%, and finally 97.7%), consistent with the fact that there are only two interstitial galleries and hence only Stage I and Stage II are meaningfully defined. Fig. 3b shows the low potential region of the potential vs. time trace in which the distinct potential plateaus can be seen more clearly.

The changes in optical transmission in our planar nano-battery allow a direct observation of the lithiation process on an individual ultrathin graphite sheet with excellent spatial and temporal resolution. As shown in Fig. 3c-g, a clear lithiated ($LiC_{36}$) and dilute stage ($LiC_{72}$) interface (i.e. a lithiation front) is observed within 100 seconds, and the lithiated area becomes more transparent and the $LiC_{36}$ area increases linearly with time (Fig. 3h). This agrees well with our electrochemical testing scheme with a constant current charge/discharge process. Thus our integrated system provides a powerful tool to investigate the intrinsic lithiation kinetics in the two-phase reaction at the nanoscale[29].

**Thickness and wavelength dependence.** We next consider the layer-number and wavelength dependence of the *in situ* optical transmission of individual ultrathin graphite sheets. In Fig. 4a-c, the wavelength dependence of the transmittance is shown in the visible range from 400 nm to 800 nm for samples of various thicknesses. For pristine ultrathin graphite (Fig. 4a), the transmittance is weakly dependent on the wavelength, consistent with previous reports[6]; the absorption by ultrathin graphite is approximately $n\pi\alpha$ ($\pi\alpha$ = 2.3%), where $\alpha$ is the fine structure constant and $N$ the number of graphene layers. This absorption results from interband transitions in the Dirac spectrum of graphene, which give a nearly constant optical conductivity $\sigma \approx \pi e^2/2h$ where $e$ is the elemental charge and $h$ Planck's constant. For $LiC_{12}$ stage (Fig. 4b), the transmittance depends more strongly on wavelength, increasing the most for longest wavelengths. For the $LiC_6$ stage (Fig. 4c), the wavelength dependent transmittance shows a maximum around 500 nm, and the transmittance of $LiC_6$ is still higher compared to pristine ultrathin graphite. Interestingly, the transmittance of $LiC_6$ still increases compared to $LiC_{12}$ for wavelengths well below the maximum, while above the transmittance maximum the $LiC_6$ transmittance decreases compared to $LiC_{12}$.

Fig. 4d and e show the optical transmittance for pristine and intercalated ultrathin graphite sheets as a function of layer number. For both 550 nm and 800 nm, a clear increase of the optical transmittance is seen after intercalation, both for $LiC_{12}$ and $LiC_6$. For 550 nm wavelength (Fig. 4d), the transmittance increases monotonically with Li concentration; for $LiC_6$ vs. pristine ultrathin graphite, the transmittance increase can be as high as 55% (for a sample of 60-80 layers). For 800 nm wavelength (Fig. 4e), the transmittance for all measured thicknesses first increases ($LiC_{12}$) and then decreases ($LiC_6$); for $LiC_{12}$ vs. pristine ultrathin graphite, the transmittance has an increase up to twofold (for a sample more than 100 layers). We also observed that in thicker sheets (insets of Fig. 4d and e), the transmittance of $LiC_6$ increases less and starts to approach the value of its pristine state at both wavelengths of 550 nm and 800 nm, i.e. at 550 nm for 90-layer thickness the transmittance at $LiC_6$ state starts to approach the value of $LiC_{12}$ state and approaches the value of pristine state at about 150 layers, and for 800 nm wavelength the transmittance of $LiC_6$ becomes lower than the value of pristine state with layer number greater than 60.

**Drude and interband contribution.** The changes in optical transmittance in the visible range described above can be qualitatively understood as follows. The result of the Li intercalation is electron doping due to the lowest electrochemical potential of Li metal. As shown in Fig, 4f, intercalation of Li heavily dopes the ultrathin graphite, shifting the Fermi level up. The magnitude of Fermi level shift is associated with the carrier density, which increases monotonically with lithium concentration. The doping concentration is as high as ~$6 \times 10^{14}$ $cm^{-2}$/layer for $LiC_6$, corresponding to $E_F \approx 1.5$ eV[30], higher than the highest doping that can be achieved in graphene with electrolytic gating[24]. The increase in Fermi energy leads to the suppression of interband optical transitions for photon energies $\omega < 2E_F$, thus decreasing the optical conductivity and increasing the transmission. As doping increases, however, intraband (Drude) absorption by free carriers becomes important, decreasing the transmission for $\hbar\omega < \hbar/\tau$ (where $\tau$ is the carrier relaxation time) due to the electron – longitudinal optical (LO) phonon interaction. Thus we expect that the transmission of ultrathin graphite is enhanced upon doping for a window of photon energies $\hbar/\tau < \hbar\omega < 2E_F$.

This phenomenon has been observed previously in gated monolayer and FLG samples, where doping levels were much lower than explored here and the window occurred in the infrared[31]. A decrease in the absorption coefficient is also observed in the graphite intercalation compound (GIC) by Hennig et al[19]. In Li-intercalated graphite, the window manifests as a minimum in reflectivity occurring near 740 nm for bulk $LiC_{12}$ and 440 nm for bulk $LiC_6$[32]. The reflectivity minimum previously observed for bulk $LiC_6$ corresponds reasonably well to our observation of a transmission maximum near 500 nm (Fig. 4c). For $LiC_{12}$ the transmission maximum may occur at a longer wavelength than our experiment accesses, and we observe only an enhancement of long-wavelength transmission. Thus we conclude that the overall reduction in interband transitions by Pauli blocking is responsible for the transmission increase, and the higher Drude conductivity of more strongly doped $LiC_6$ is responsible for the observed reduction in transmission at long wavelength and the non-monotonic doping dependence of transmission at these wavelengths.

Further insight into these results is gained by examining the optical transmittance in terms of the optical conductivity. On a substrate with refractive index $n$, the transmittance of ultrathin graphite with optical (sheet) conductivity $\sigma_{opt} = \sigma_1 + i\sigma_2$, relative to that of the bare substrate, can be expressed as[33]:

$$T = \frac{1}{\left|1+\frac{Z_0\,\sigma_{opt}}{1+n}\right|^2} \quad (1)$$

where $Z_0$ is the free space impedance. Li *et al.*[31] and Stauber *et al.*[44] reported the optical conductivities of doped monolayer graphene in the IR range. For $\hbar\omega < 2E_F$, i.e. below the Pauli blocking edge, $\sigma_1 \gg \sigma_2$; $\sigma_1$ is large due to LO phonon emission and $\sigma_2$ passes through zero near the plasma edge ($\varepsilon_1 = 0$) (also see the Supplementary Information) so that the transmission reaches a maximum. Therefore near the transmission maximum $\sigma_{opt} \sim \sigma_1$.

We then modeled the optical transmittance based on optical conductivity with a Drude contribution from the free carriers and an interband contribution that turns on for $\hbar\omega > 2E_F$ due to Pauli unblocking, i.e., $\sigma_{opt}(\omega) = \sigma_D + \sigma_{ib}$. The conductivity is modeled as $N$ layers of graphene. The Drude sheet conductivity can be written as[2] $\sigma_D = \frac{n_{2D}e^2 N}{m(\gamma - i\omega)} = \frac{e^2 E_F N}{\pi\hbar(\gamma - i\omega)}$, where $n_{2D}$ is the carrier density per layer, $m$ the effective mass, $\gamma = 1/\tau$ is the carrier relaxation rate, and $N$ is the number of layers. The thermal broadening of the Pauli blocking leads to[34] $\text{Re } \sigma_{ib} = \frac{\pi e^2 N}{2h}\left[\tanh\left(\frac{2E_F + \hbar\omega}{4kT}\right) + \tanh\left(\frac{2E_F - \hbar\omega}{4kT}\right)\right]$, and the imaginary part of $\sigma_{ib}$ is obtained from Kramers-Kronig relation. The details of the modeling are presented in the Supplementary Information. Fig. 4g shows a schematic of $\sigma_1 / N\sigma_0$ vs. the photon energy from the model for $N$-layer ultrathin graphite before and after Li intercalation. $\sigma_1$ in the visible range significantly decreases upon Li intercalation, which leads to a large increase in the optical transmittance. The modeled transmittance of both 8-layer and 83-layer ultrathin graphite (Fig. 4h, solid curves) closely resemble the corresponding experimental data, and the sharpening of the transmission maximum for thicker films is a consequence of the plasma edge as discussed in the Supplementary Information.

**Electrical properties.** In order to understand the prospects for highly transparent Li-intercalated ultrathin graphite for conducting transparent electrode applications, we adapted our planar nano-battery setup for *in situ* conductivity measurements of ultrathin graphite during electrochemical cycling. We transferred ultrathin graphite onto pre-deposited electrical contacts in a Hall-bar arrangement (Fig. 5a inset, see also Supplementary Information). Fig. 5a shows the room temperature sheet resistance $R_s$ for ultrathin graphite samples with different thickness before intercalation as well as intercalated to $LiC_{12}$ and $LiC_6$. As expected, all intercalated ultrathin graphite samples invariably exhibit a lower resistivity compared to their pristine state (also see Supplementary Information). Note that $R_S$ measured on both stage I and II is inversely proportional to the sample thickness (before intercalation) as indicated by the dashed lines. Considering the expansion of the graphite-layer spacing during Li-intercalation[35] we can estimate that $\rho(LiC_6) \sim 3.1 \times 10^{-6}$ $\Omega \cdot$cm and $\rho(LiC_{12}) \sim 1.4 \times 10^{-5}$ $\Omega \cdot$cm. The intrinsic limit of the conductivity for doped graphene at room temperature is set by electron-acoustic phonon scattering and is approximately $\sigma_{dc,phonon} = 33$ mS per layer[26,36] for Fermi energies in the linear portion of the band structure, while we observe a DC sheet conductivity $\sigma_{dc} \approx 11$ mS per layer in $LiC_6$. At the high doping levels present in $LiC_6$, we expect significant band curvature and reduction in the Fermi velocity, likely reducing the limiting conductivity. Additionally, disorder may play a role. Thus our approach within a factor of ~3 to the limiting conductivity value for the graphene Dirac band is impressive.

In order to elucidate the type and density of charge carriers we investigated the Hall resistance at perpendicular magnetic fields. The linear $R_{xy}(B)$ curves with negative slope (Fig. 5b) indicate that charge carriers are electrons for a 4 nm-thick FLG device after Li-intercalation. The carrier density $n_H$ is readily determined by a measurement of the Hall coefficient $R_H = R_{xy}/B$, where $R_H$ is related to $n_H$ by $n_H = 1/eR_H$. With the information of expanded thickness of lithium intercalated ultrathin graphite, our measurements reveal that bulk $n_H$ ranges from $3 \times 10^{21}$ to $7 \times 10^{21}$ cm$^{-3}$ for $LiC_{12}$ and from $1.5 \times 10^{22}$ to $3.5 \times 10^{22}$ cm$^{-3}$ for $LiC_6$, with no observable dependence on sample thickness, as shown in Fig. 5c. These values compare reasonably well with the full ionization values of $1.7 \times 10^{22}$ cm$^{-3}$ for $LiC_6$ and $9.0 \times 10^{21}$ cm$^{-3}$ for $LiC_{12}$, which are indicated in Fig. 5c as guidelines. Fig. 5d shows the temperature dependence of the sheet resistance. $R_s(T)$ is metallic, i.e. $R_S$ decreases with decreasing $T$, for samples at $LiC_6$ state, while $LiC_{12}$ exhibits a moderate temperature dependence, and pristine samples always exhibited weakly non-metallic behavior consistent with previous studies[3]. The strong decrease in $R_S$ with lowering $T$ for $LiC_6$ is consistent with phonon-limited conduction and further corroborates that we have approached the phonon-limited conductivity in Li-intercalated ultrathin graphite [21].

**Transparent electrode performance.** In the race to find better transparent electrodes researchers have investigated numerous candidate materials.[25,37-40] Fig. 6 shows the transmittance vs. sheet resistance of Li-intercalated ultrathin graphite as well as other high-performance transparent conducting materials, including other carbon-based materials,[15,41] and the best commercial indium-tin-oxide (ITO) electrodes.[39,42] In previous doped-graphene studies an improvement in the electrical conductivity was observed; however, little or no change of transmittance in the visible range was obtained.[9,15] When $\sigma_1 \gg \sigma_2$ and $n = 1$, to compare the performance of a freestanding film in vacuum, Equation (1) becomes:

$$T = \frac{1}{\left(1 + \frac{[188\ \Omega]}{R_S} \frac{\sigma_{opt}}{\sigma_{dc}}\right)^2} \quad (2)$$

where the sheet resistance is $R_s = 1/\sigma_{dc}$. Thus at a given sheet resistance, the transmission is determined by the ratio $\sigma_{dc}/\sigma_{opt}$ which can be used as the Figure of Merit (FOM) to characterize the performance of a transparent conductor.

As shown in Fig. 6a we fit the data for our Li-intercalated FLG devices to Eqn. 1 using $\sigma_{dc}/\sigma_{opt}$ as a fitting parameter, and fitting result gives $\sigma_{dc}/\sigma_{opt} = 920$. For the best sample at $LiC_6$ state we measured transmittance of 91.7 % and 3.0 $\Omega$ per sq, obtaining $\sigma_{dc}/\sigma_{opt} = 1400$. Fig. 6b shows the best measured $\sigma_{dc}/\sigma_{opt}$ for the material systems shown in Fig. 6a; $\sigma_{dc}/\sigma_{opt}$ for our $LiC_6$ exceeds that of FeCl$_3$ intercalated FLG ($\sigma_{dc}/\sigma_{opt}$ = 235)[15] and the best commercial transparent electrode ITO ($\sigma_{dc}/\sigma_{opt}$ = 118)[39]. In fact, $\sigma_{dc}/\sigma_{opt}$ of few-layer LiC$_6$ exceeds that for all other carbon based materials, and as far as we can determine is the highest for any uniform thin film. Higher transparency at a given conductivity has only been achieved in inhomogenous conductors such as metal nanowire networks[43], which may not be suitable for many applications. It also exceeds the intrinsic limit for doped graphene[9,37] previously expected ignoring the increased transparency due to Pauli blocking. The sheet resistance and transparency easily meet the need for optoelectronic device applications where 90% and 10 Ω per sq is required. Thus we expect that electrochemically intercalated FLG is promising for applications where the highest DC conductivity at a given optical transparency is needed.

To demonstrate the feasibility of ultrathin graphite as transparent electrode for industrial applications, we successfully fabricated stable millimeter-scale devices using encapsulated commercially-obtained chemical vapor deposition-grown (CVD) thin graphite (see Methods and Supplementary Information). Comparison of two 40 nm and 80 nm thick devices before and after Li intercalation are shown in Fig. 7a, b. Transmittance spectra are also characterized before and after Li intercalation, as shown in Fig. 7c, which are very similar to the results of single exfoliated ultrathin graphite sheets. Changes in the sheet resistance was also measured for 40-nm-thick devices by the Van der Pauw method (see Supplementary Information). The sheet resistances of three different devices dropped drastically upon complete lithiation from 35.4, 47.7, 57.0 Ω per sq (graphite) to 3.0, 3.9, 1.7 Ω per sq (LiC$_6$), respectively. The 1.7 Ω per sq correspond to a single layer sheet resistance of 200 Ω per sq, only 2.3 times larger than the single flake value (87.3 Ω per sq) from our experiment, leads to a FOM of 400.

**Discussion**

We discuss the ultimate limits to conductivity and transparency of doped graphene-based systems. Previous studies[25] of graphene as a transparent conductor have ignored changes in the optical conductivity, assuming it remains limited by interband transitions and is fixed at $\sigma_{opt} = \sigma_{ib} \approx N\pi e^2/2h = N\sigma_0$, with $N$ the number of layers, as discussed above. For phonon-limited conduction at room temperature $\sigma_{dc} = N\sigma_{dc,phonon}$ where $\sigma_{dc,phonon}$ = 33 mS[26,36]. This predicts a maximum value of $\sigma_{dc}/\sigma_{opt} \approx 550$, while our intercalated ultrathin graphite significantly exceeds this value. However, as noted previously, below the Pauli blocking edge $\sigma_{opt}$ is the free carrier Drude conductivity that can be smaller than $\sigma_{ib}$. Optical measurements on doped monolayer graphene gives $f \equiv \sigma_1/\sigma_0 \approx 0.3$ below the interband edge for $E_F \approx 0.3$ eV [31]. Assuming this value of $f$ implies $\sigma_{dc}/\sigma_{opt} \approx 1800$ in reasonable agreement with our best observation; however, there are no experimental results on the magnitude of $\sigma_1$ and hence $f$ for the $E_F \approx 1.5$ eV conditions of our intercalated graphene. Theory predicts that the Drude optical conductivity for frequencies above the LO phonon frequency is limited by the electron - LO phonon relaxation rate, $\gamma = 1/\tau_{LO}$ [44]. At the higher $E_F$ of our experiments the electron phonon scattering rate will be stronger because of the larger electronic density of states ($\sim E_F$), but the high frequency Drude conductivity falls off as $\sigma_1 \sim E_F\, \gamma/(\gamma^2+\omega^2) \sim E_F\, \gamma/\omega^2$. This suggests $\sigma_1$ and hence $f$ at $\omega \leq 2E_F$ are approximately independent of $E_F$ so that $f \equiv \sigma_1/\sigma_0 \approx 0.3$ may also be valid at $E_F \approx 1.5$ eV, and our estimate of the intrinsic limit of $\sigma_{dc}/\sigma_{opt} \approx 1800$ is reasonable. Thus we believe that our real devices approach the ultimate limits of transparency at a given conductivity for the doped graphene system.

In summary, we have designed a methodology *via* a planar nano-battery for *in situ* study of the electrical and optical properties of individual ultrathin graphite sheets during electrochemical intercalation and de-intercalation. Metallic-like temperature dependent transport is observed in Li-intercalated ultrathin graphite with conductivities approaching the acoustic-phonon limit at room temperature and is comparable to good metals. Due to the unusual band structure of graphene, Li-intercalation can simultaneously increase the DC electrical conductivity and increase optical transmission in the visible, allowing Li-intercalated FLG to achieve an unprecedented FOM $\sigma_{dc}/\sigma_{opt}$ = 920, significantly higher than any other material and approaching the ultimate limit expected for doped graphene systems. Our technique will allow similar studies to be carried out in other 2D materials. Furthermore, the methodology reported in this study can be applied to *in situ* investigations of the intercalation process with good spatial and temporal resolution in materials for electrochemical energy storage applications.

## Methods

**Devices fabrication for *in situ* electrochemical and optical measurements.** Pristine ultrathin graphite sheets from monolayer to 50nm (~150 layers) are first obtained by mechanical exfoliation of Kish graphite onto 0.2 mm thick glass substrate (Fisher Scientific), followed by deposition of electrical contacts (50 nm copper) on top of selected ultrathin graphite sheets using a shadow mask technique in electron beam evaporator. The device is then transferred into a glove box filled with argon gas, and a small lithium pellet is deposited onto an isolated electrical contact, followed by the addition of a small amount of electrolyte ($LiPF_6$ in EC:DEC w:w = 1:1) to cover the region with both ultrathin graphite and lithium pellet. At last the center region with electrolyte/lithium/ultrathin graphite is covered by another piece of 0.2 mm thick glass and sealed by PDMS, as shown in Fig. 1c.

**Device fabrication for electrical transport measurements.** During the Li intercalation process the volume of ultrathin graphite gradually expands because of the insertion of lithium atoms. The layer spacing of LiC$_6$ is ~10% larger than that of pristine graphite[35]. Therefore, the narrow metal electrodes fabricated by the normal method of thermal evaporation on top of ultrathin graphite usually crack after intercalation. Here we use a lithography-free fabrication method shown in Fig. 1d-f. The 50 nm thick copper Hall-bar/lithium-contact electrodes are pre-patterned on a blank glass wafer. A uniform exfoliated ultrathin graphite sheet is then transferred onto the top of the electrodes aligned by a micro-manipulator. The rest of the device fabrication is the same as described above. Using this method, ultrathin graphite sheets are attached to the top of the electrodes and can expand freely during Li-intercalation.

**Fabrication of large scale CVD graphene devices.** CVD ultrathin graphite on Nickel foil (2" × 2") is obtained from Graphene Supermarket and cut into 1.5 cm×1.5 cm pieces. A solution based (1M FeCl$_3$ in DI water as etchant, Sigma Aldrich) etching/transfer method is then carried out to transfer ultrathin graphite onto transparent substrates (e.g. glass and PET). A gel electrolyte film is prepared by mixing P (VDF-HFP)/Acetone/DI water (w:w:w = 1:19:1, Sigma Aldrich) as a mixed solution. The electrolyte is then drop-cast on glass and ready for use after drying in a vacuum oven (MTI corp.). A sandwiched device structure of glass/ultrathin graphite/gel electrolyte/PET is assembled in an argon-filled glove box. The transmittance of the device is measured by a UV-vis spectrometer (PerkinElmer Lambda 35).

*In situ* **optical transmission measurements.** A system based on a transmission optical microscope (Nikon Eclipse Ti-U) combined with a charge-coupled device (CCD) camera is used to acquire transmission data. Microscope objectives with 5× and 20× magnification are used depending on the size of the sample. A beam of light at particular wavelength from a monochromator (Spex 500M, 0.2 nm bandwidth) passes through the transmission optical microscope and is then projected onto the 1392 × 1040 lines of the grey CCD camera. Intensity is then extracted from images taken by CCD camera and is normalized to the signal obtained through a region of bare substrate close to the sample to give the transmission at that wavelength. Such analysis results in a weak overestimate of transmittance for ultrathin graphite embedded in an electrolyte solution, however if we focus on the highly transparent samples (more than 90%) the difference is negligible (~1%).

**Electrochemical control.** An electrochemical workstation (Biologic SP-150) is used to control charge/discharge of the Li-ultrathin graphite nanobattery and measure the time-dependent potential on intercalation (lithiation) and de-intercalation (delithiation). Details are also discussed in Supplementary Information.

**Acknowledgements**

The work is supported by Energy Research Frontier Center (EFRC) funded by DOE, Office of Science, BES under Award # DESC0001160, and the U.S. ONR MURI program. This work is partially supported by NSF-CMMI Award #1300361. J. N. Munday acknowledges the startup support from University of Maryland. M.S. Fuhrer acknowledges support from an ARC Laureate Fellowship. We acknowledge the support of the Maryland Nanocenter. We thank Dr. Karen Gaskell, Joe Murray and Joe Villeneuve for technical assistance for this experiment. Also thanks to Jiaqi Dai for making Cover Art and Colin Preston for reading the manuscript and useful comments.

**Author contributions**

L. H., W. B., J. W., and M. S. F. designed the experiments; W. B., J. W. and H. C. fabricated all devices; W. B. and J. W. carried out *in situ* electrochemical and optical measurements; W. B., J. W. and D. K. carried out electrical transport measurements; W. B. and J. W. did data analysis; H.D.D. led the theoretical interpretation; X. H. and J. W. measured the Raman spectrum; J. W., D. M., W.B. and H. Z. carried out AFM measurements; Y. X. and J. M. assisted with *in situ* optical measurement; W. B., J. W., H.D.D., M.S. F. and L. H. wrote the manuscript, with input from all authors.

**Figure 1. Device schematics.** (**a-c**) Schematic of the fabrication process of planar nano-battery devices for *in situ* optoelectronic measurement. (**d-f**) Schematic of the fabrication process of devices for electrical transport measurements. See also Methods for details of device fabrication.

**Figure 2.** *In situ* **optical and electrochemical measurement with a planar nano-battery platform.** (**a-c**) Electrochemical battery tester and transmission optical microscope are integrated for *in situ* measurement of individual ultrathin graphite sheets on glass substrates. (**c**) An AFM image of a uniform ultrathin graphite sheet attached to the electrical contact. The scale bar is 10 μm. (**d-i**) Transmission optical microscope images of ultrathin graphite before intercalation and at different intercalation stages as indicated in the figure panels. A schematic of the $LiC_6$ lattice structure is also shown in (**h**). The scale bar in (**d**) is 100 μm. (**j**) Raman spectra of Li intercalated ultrathin graphite at different intercalation stages as indicated in the figure panel.

**Figure 3. Optical transmittance evolution during electrochemical Li intercalation process.** (**a**) Optical transmittance (right) and electrochemical potential (left) vs. lithiation time are plotted. (**b**) Detail of voltage profile vs. time near the intercalation plateau. (**c-g**) Optical images of an ultrathin graphite sheet at different time points during intercalation as indicated in each panel, showing a clear lithiation front (red dashed line) between $LiC_{36}$ (lighter contrast) and $LiC_{72}$ (darker contrast). The ultrathin graphite sample is about 120 layers thick and the scale bar in (**c**) is 20 μm. (**h**) Lithiated $LiC_{36}$ area vs. time extracted from images such as (**c-g**).

**Figure 4. Wavelength-dependent optical transmittance of intercalated ultrathin graphite.** (**a-c**). Transmittance as a function of wavelength for different thickness ultrathin graphite samples for pristine ultrathin graphite (**a**), LiC$_{12}$ (**b**), LiC$_6$ (**c**) stages. (**d-e**) Transmittance as a function of thickness plot for pristine ultrathin graphite, LiC$_{12}$, and LiC$_6$ at wavelengths 550 nm (**d**) and 800 nm (**e**). Insets show transmittance at same wavelength over a larger range of thicknesses. (**f**) Schematic of doped graphene bandstructure illustrating suppression of optical transitions due to Pauli exclusion principle. (**g-h**) Modeled results for the real part of optical conductivity, $\sigma_1/N\sigma_0$ (**g**), and transmittance (**h**) of pristine ultrathin graphite (dashed line) and intercalated LiC$_6$ (solid line). For the model we assume $E_F$ = 1.5 eV, $n$ = 1.5, $T$ = 300K, and $1/\tau = \gamma$ = 200 cm$^{-1}$. The red and black colors in (**h**) correspond to 8 and 83-layer ultrathin graphite sheets, respectively. Experimental data of 8 and 83-layer shown in (**c**) are also plotted in (**h**) for comparison with the model.

**Figure 5. Transport measurement of Li intercalated ultrathin graphite sheets.** (**a**) Resistivity *vs.* thickness for ultrathin graphite sheets with different thickness. Data for pristine (red) and two lithiated stages (LiC$_{12}$ and LiC$_6$ indicated as green and blue) are shown. Inset: An optical image of an ultrathin graphite device with Hall bar geometry before intercalation. The scale bar is 10 μm. (**b**) Hall resistance of a 4-nm-thick pristine FLG sheet and its LiC$_{12}$ and LiC$_6$ states as a function of magnetic field. (**c**) Carrier density calculated from Hall measurement as a function of ultrathin graphite thickness. (**d**) Temperature dependent sheet resistance for two ultrathin graphite samples. Blue, red and green colors indicate pristine, LiC$_{12}$ and LiC$_6$ stages, respectively.

**Figure 6. Optoelectronic properties of intercalated ultrathin graphite sheets and comparison with other materials.** (**a**) Transmittance at 550 nm vs. sheet resistance for our LiC$_6$ FLG, and

other high-performance carbon-based transparent conducting materials FeCl$_3$-doped graphene[15], acid-doped graphene[9], and carbon nanotube (CNT) films[37], as well as indium tin oxide (ITO)[39]. The red solid line is a fit with equation 3 with $\sigma_{dc}/\sigma_{opt} = 920 \pm 100$. (**b**) Figure of merit ($\sigma_{dc}/\sigma_{opt}$) for various materials. A higher value for $\sigma_{dc}/\sigma_{opt}$ leads to better performance in transparent conductor.

**Figure 7**. **Demonstration of encapsulated large-area transparent electrode (a-b)** Photographs of 40- and 80-nm thick CVD grown thin graphite before and after full Li intercalation. (**c**) Corresponding transmittance spectra of the two devices before and after full intercalation.

**Figure 1**

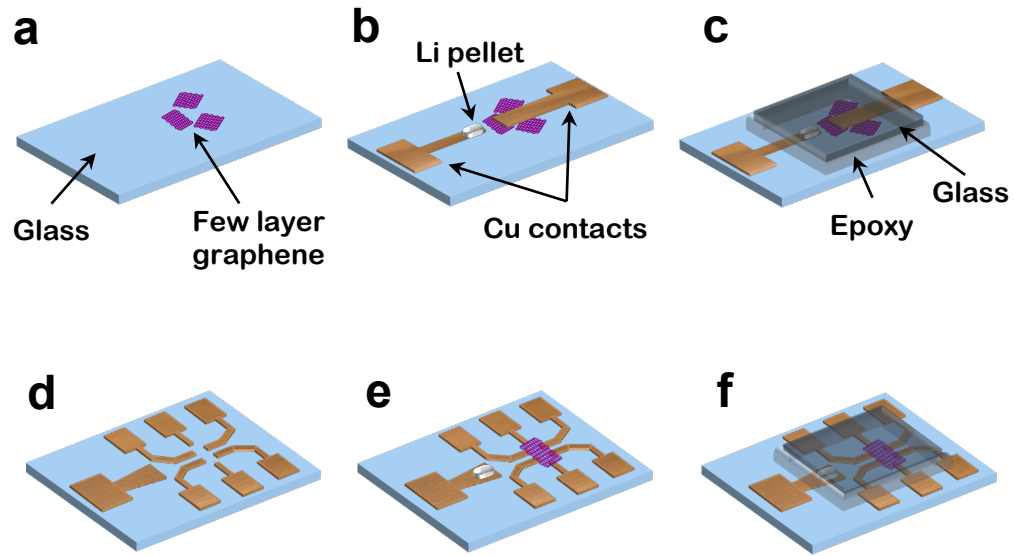

# Figure 2

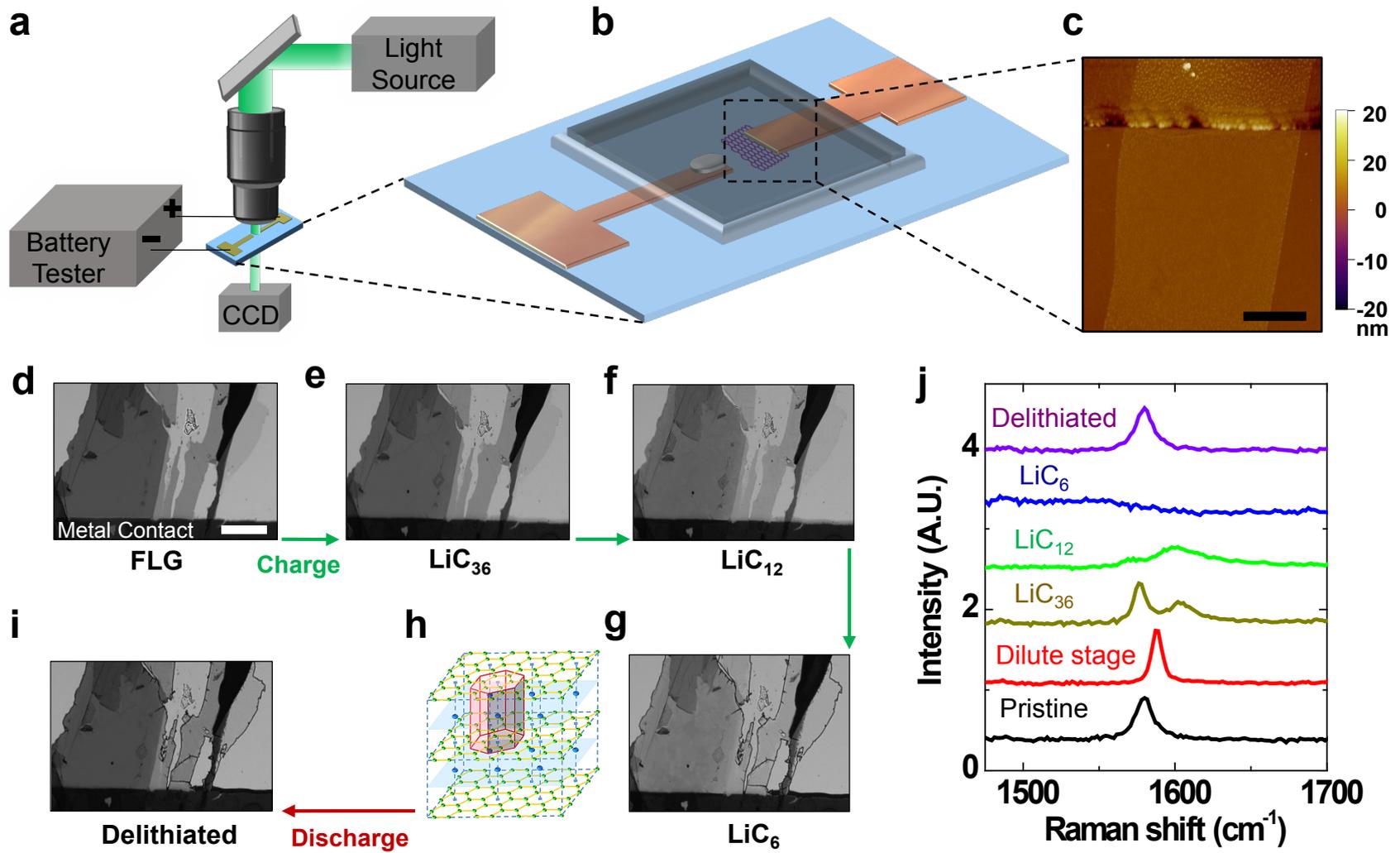

# Figure 3

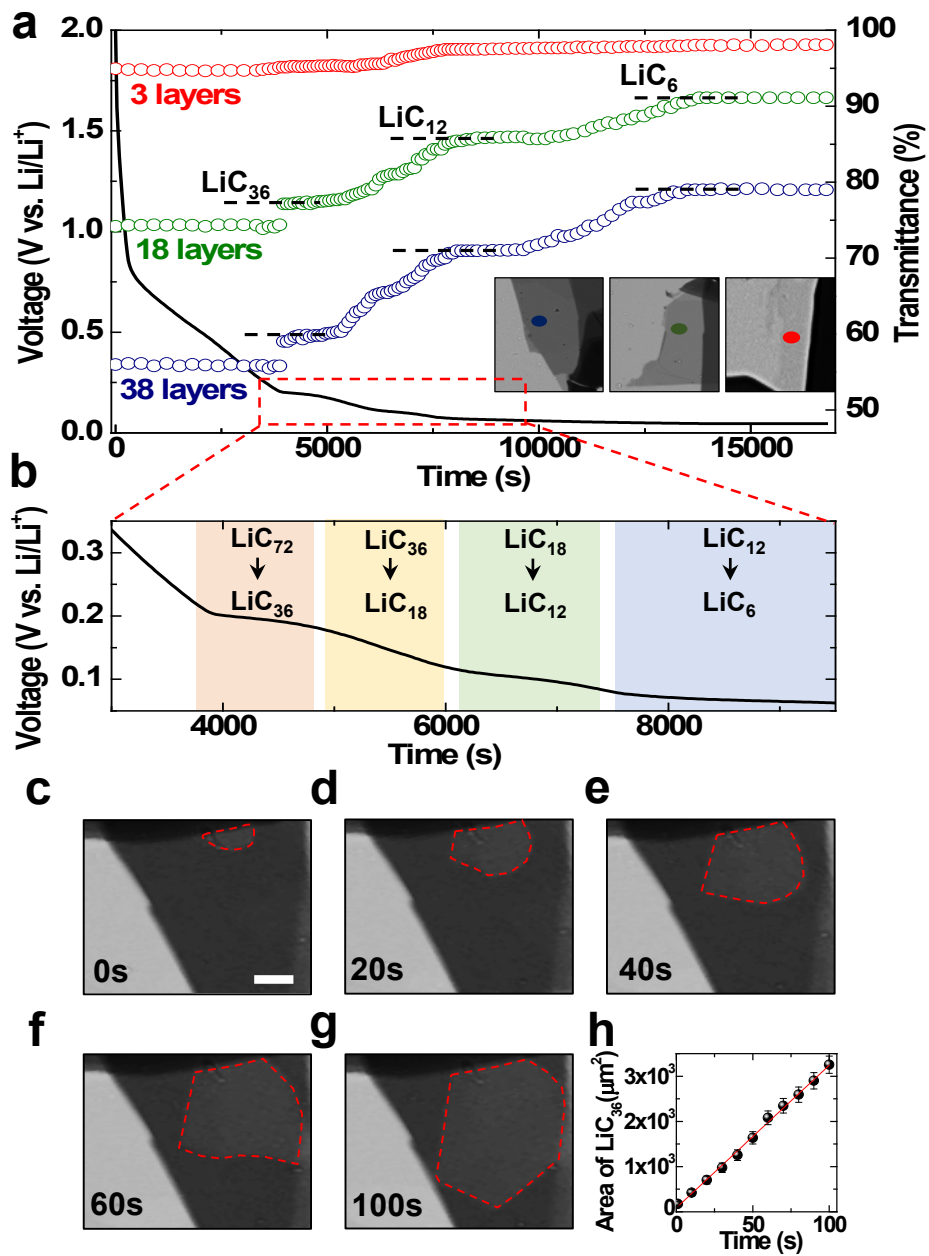

# Figure 4

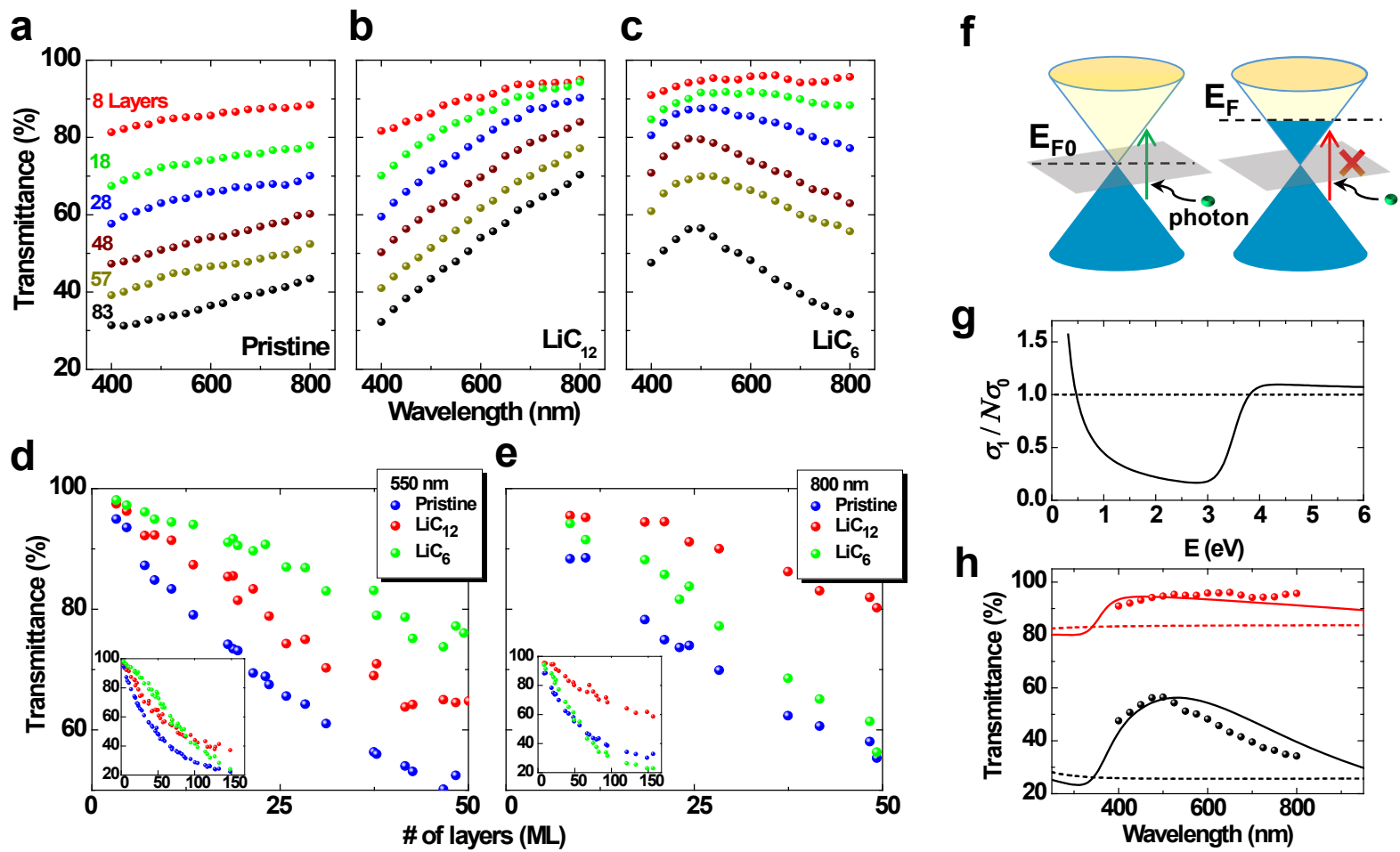

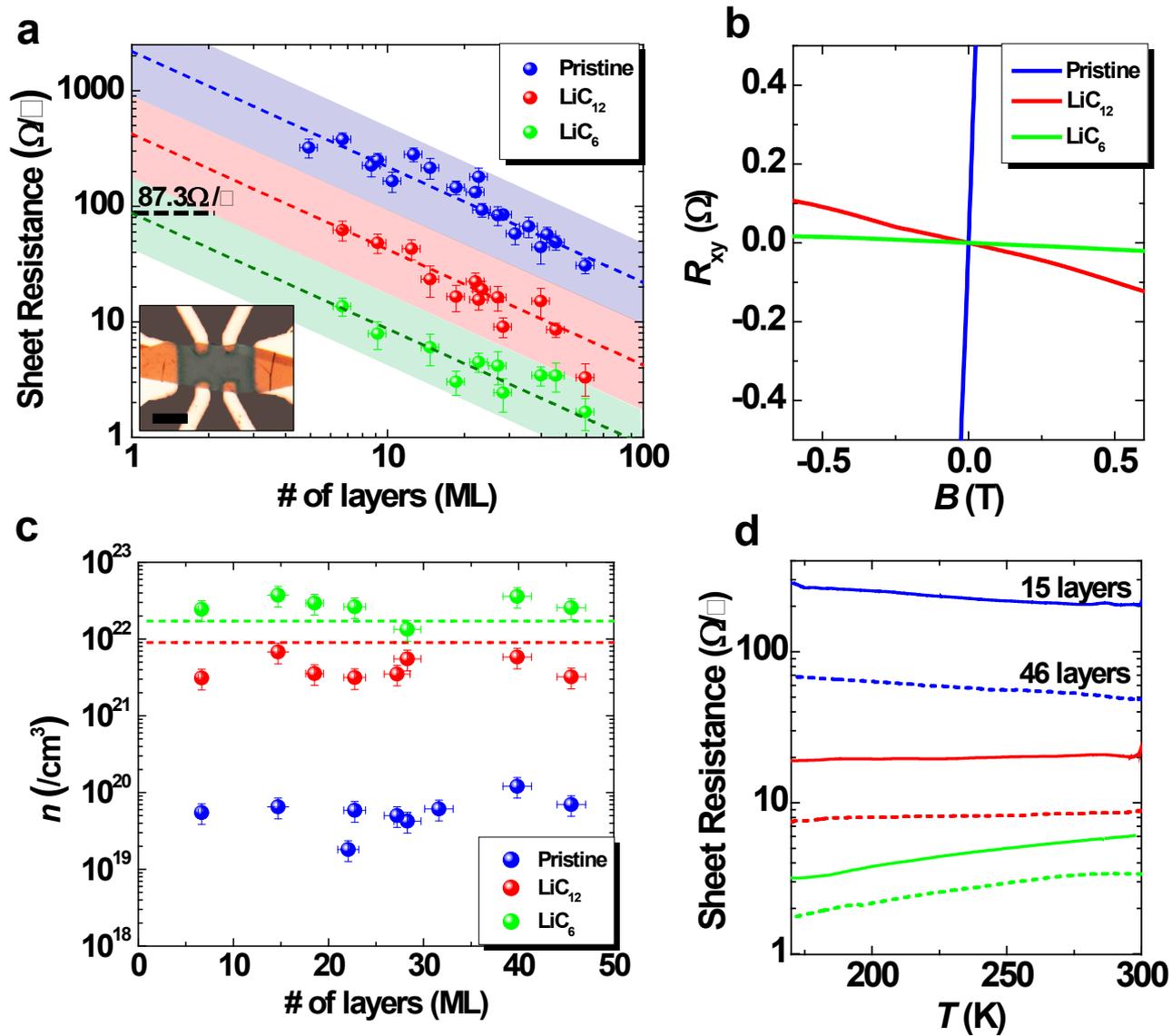

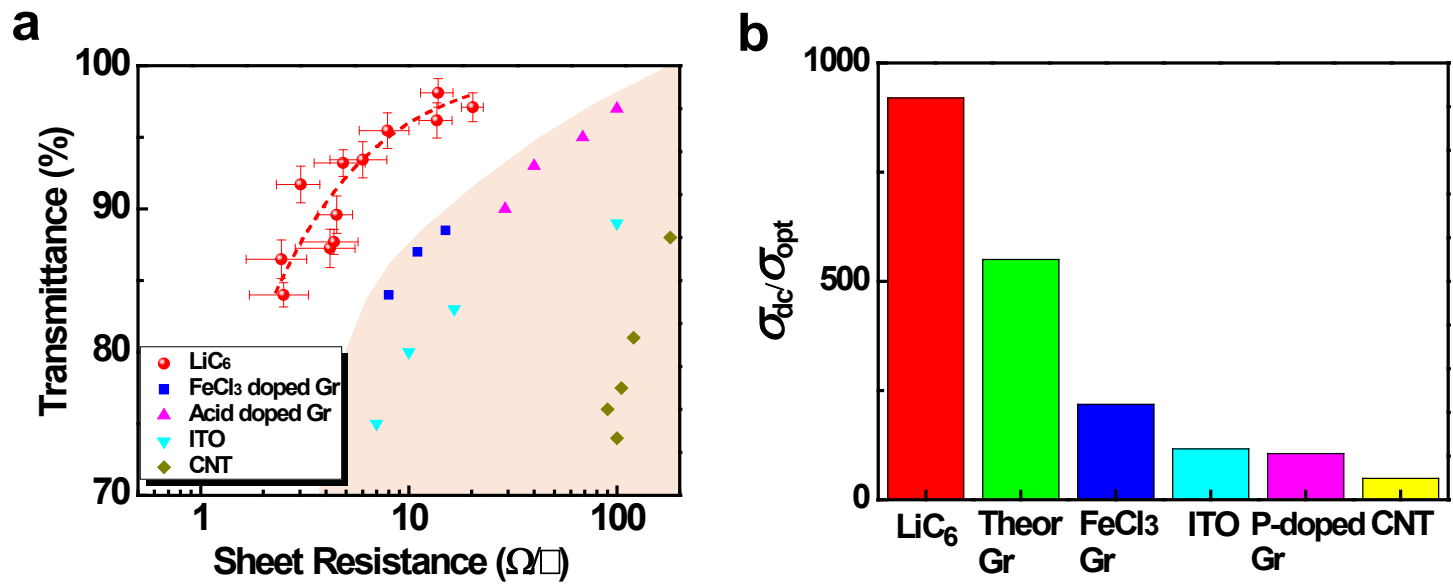

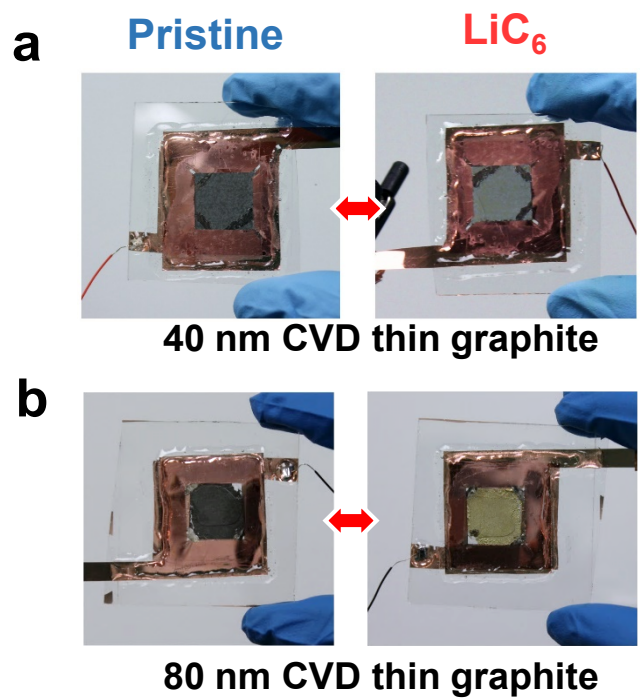
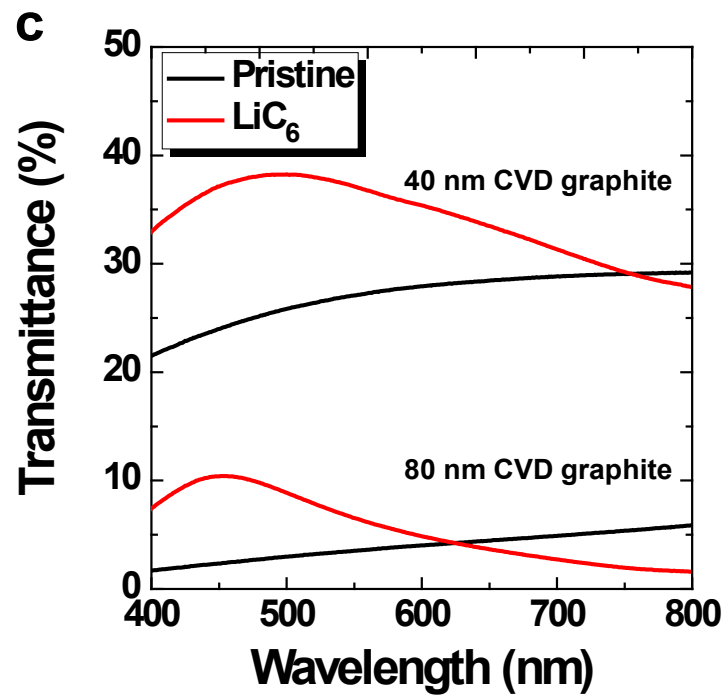

# Supplementary Information

**Supplementary Figures**

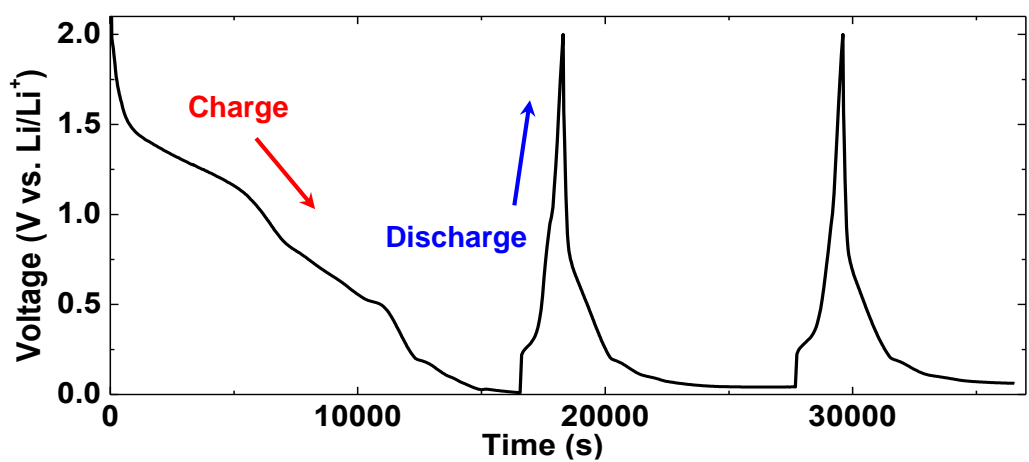

**Supplementary Figure 1.** Typical voltage profile of Li-ultrathin graphite half-cell at first few charge and discharge cycles.

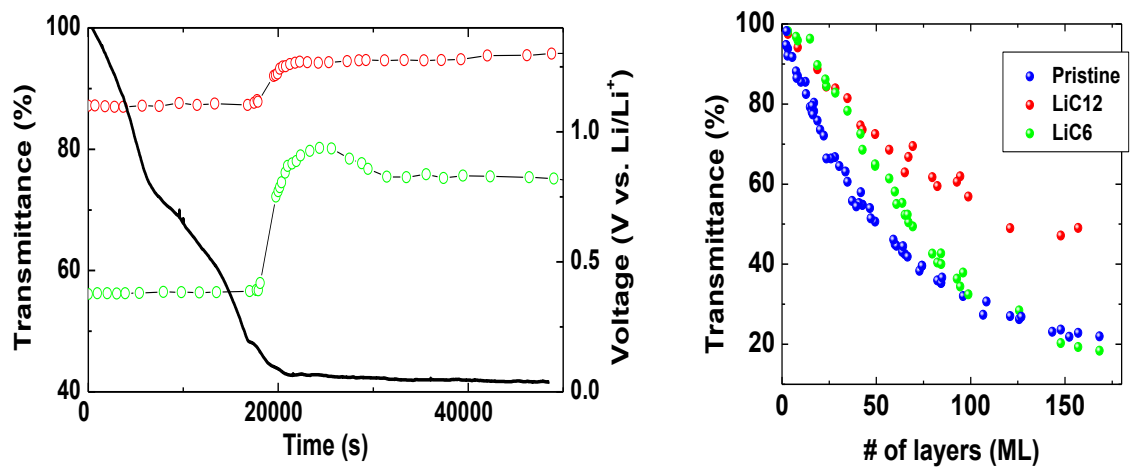

**Supplementary Figure 2.** (**a**) *In situ* white light transmittance (open circles) and electrochemical potential (solid line) *vs.* time during Li interaction for two ultrathin graphite sheets with different thickness. (**b**) White light transmittance *vs.* layer number for pristine thin graphite, $LiC_{12}$, and $LiC_6$.

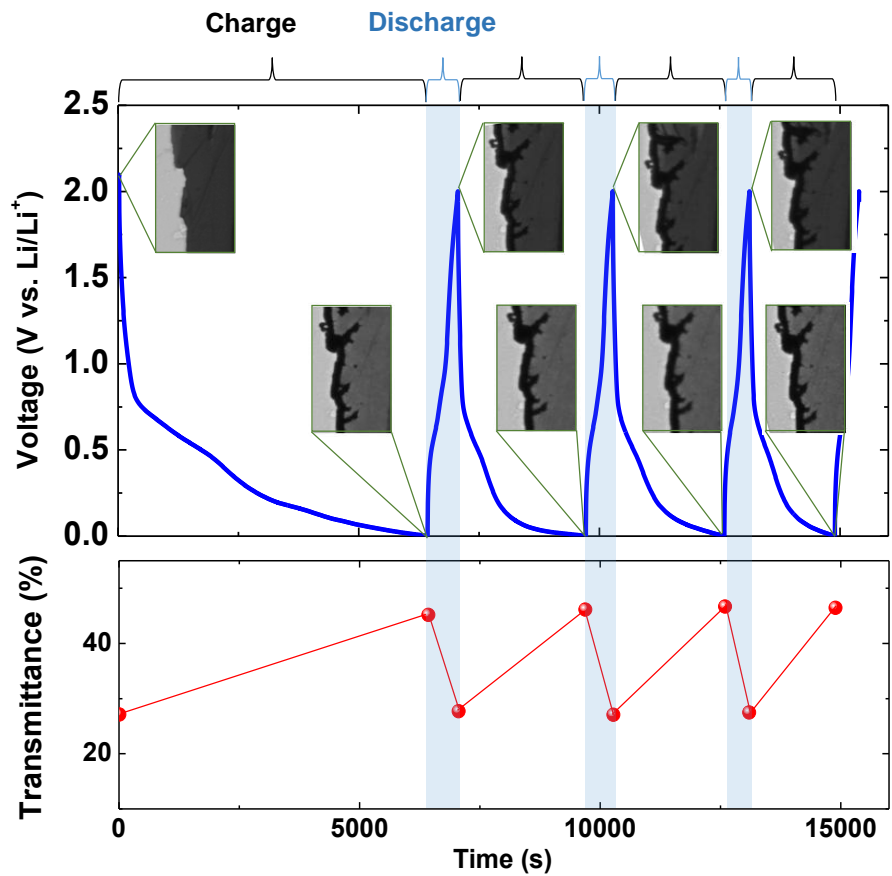

**Supplementary Figure 3.** Electrochemical potential (upper panel) and optical transmittance (at 550 nm; lower panel) of an ultrathin graphite sheet (about 100 layers) during multiple charge/discharge cycles.

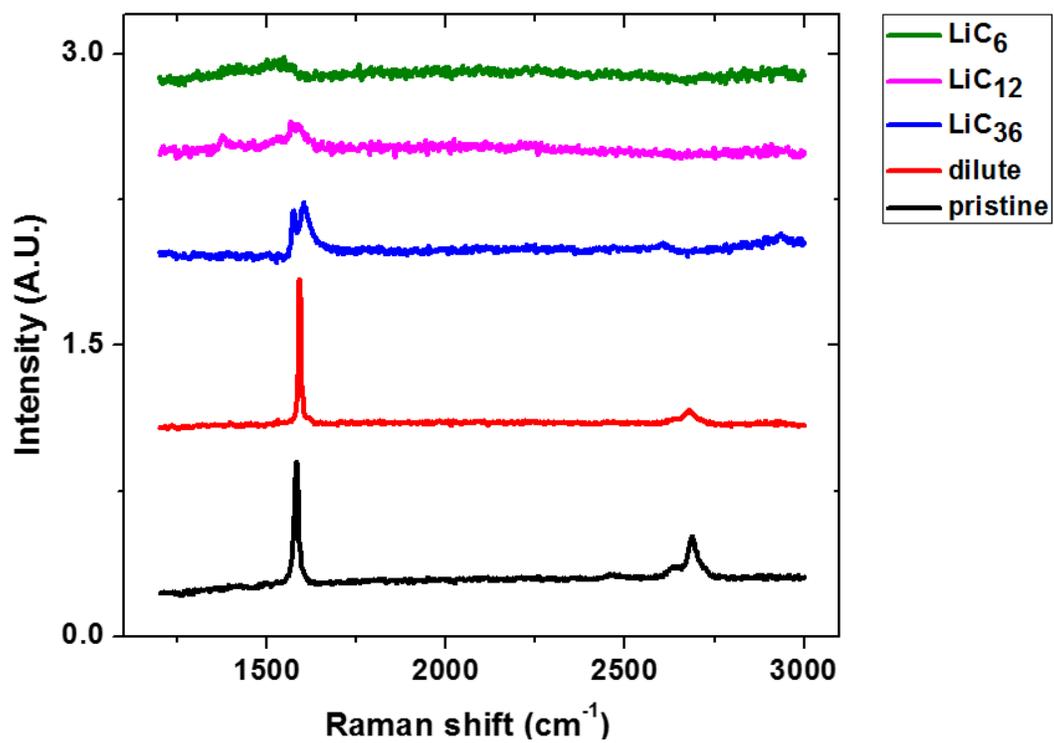

**Supplementary Figure 4.** Raman spectra of pristine and lithiated ultrathin graphite.

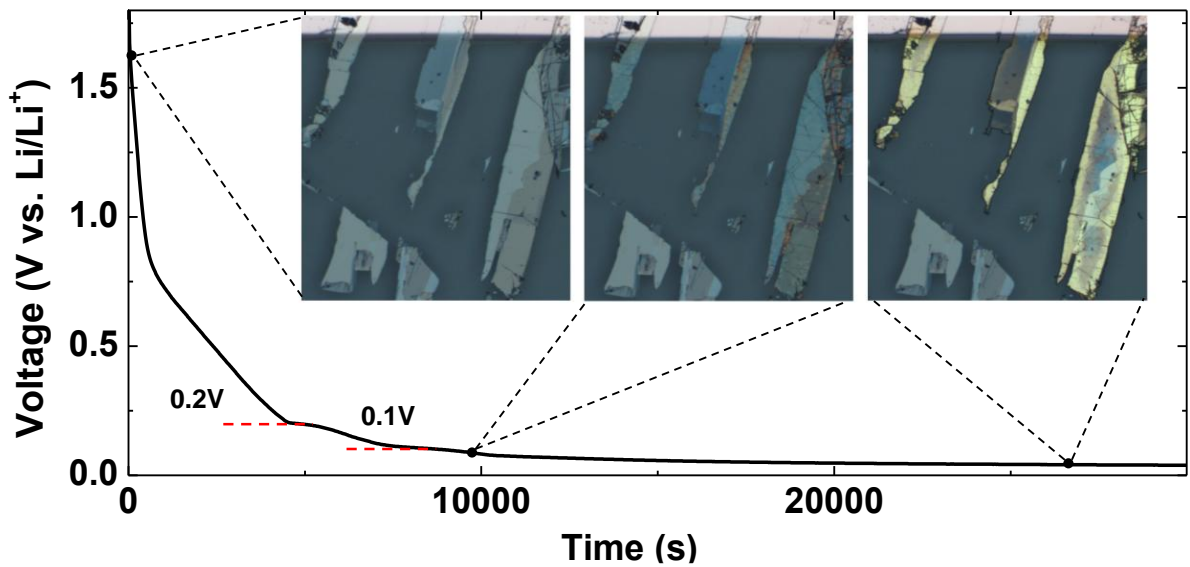

**Supplementary Figure 5**. *In situ* optical (reflective mode) and electrochemical measurement of ultrathin graphite sheets. 0.5 µA charge current is applied for voltage potential *vs*. lithiation time measurement.

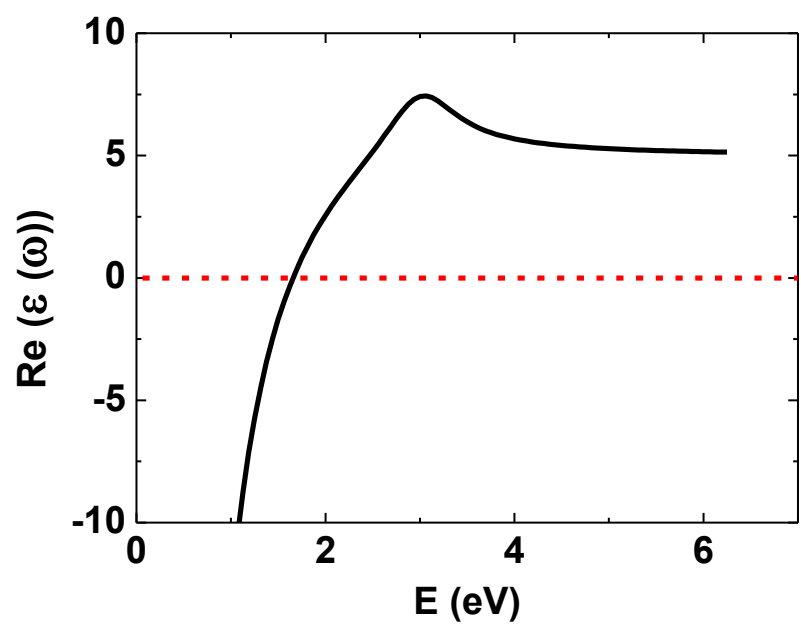

**Supplementary Figure 6.** Modeled result for the real part of dielectric constant $\varepsilon$ vs. photon energy.

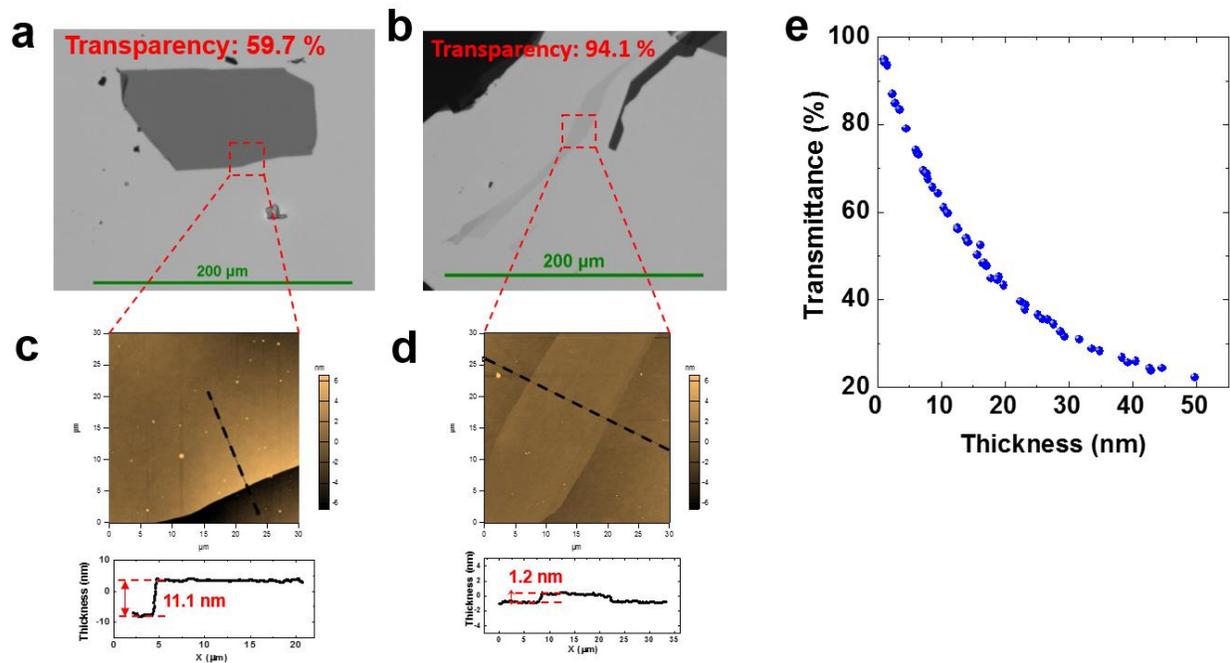

**Supplementary Figure 7. a-b** Optical transmission (550nm light source) images of uniform isolated ultrathin graphite samples deposited on glass substrate by mechanical exfoliation. **c-d** corresponding AFM images. **e**. Transmittance (550nm light source) *vs* AFM thickness for all measured flakes.

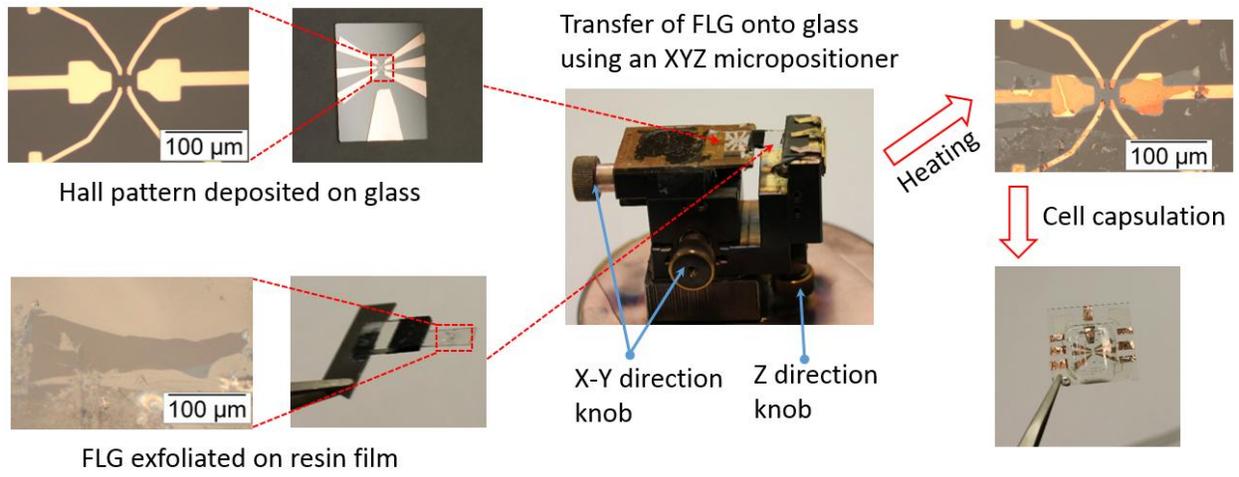

**Supplementary Figure 8.** Process flow of ultrathin graphite transfer and cell encapsulation.

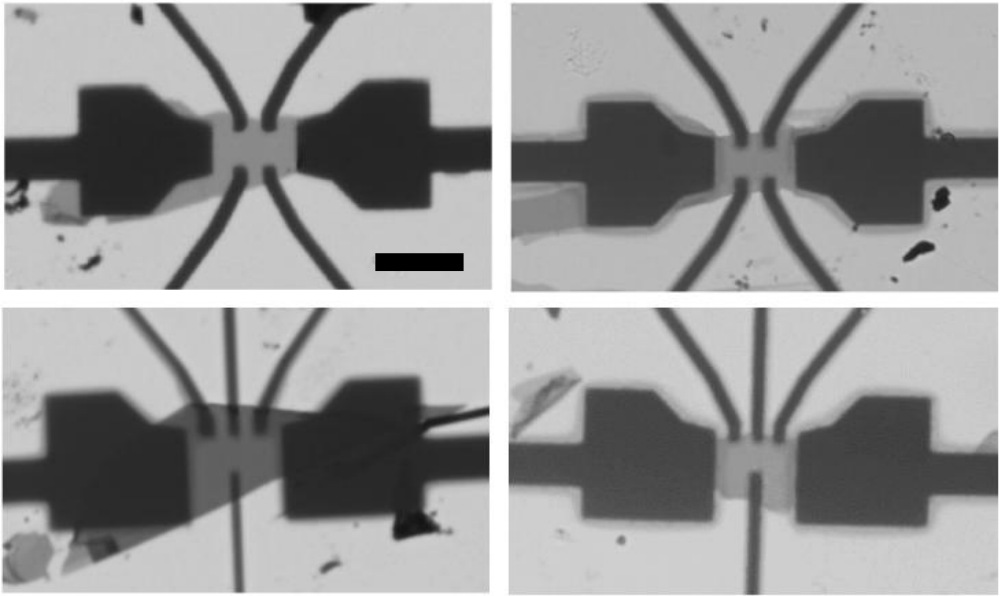

**Supplementary Figure 9.** Optical transmission images of devices with ultrathin graphite sheets transferred onto pre-fabricated Hall bar electrodes. The scale bar is 20 μm.

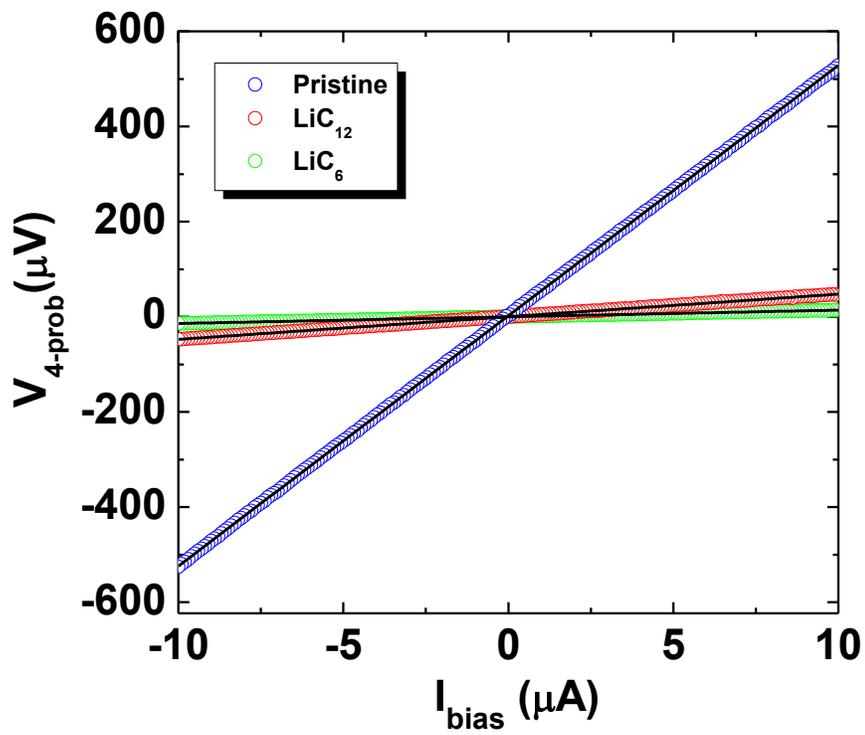

**Supplementary Figure 10.** Four-probe I-V plots for pristine and lithiated states of an ultrathin graphite sheet.

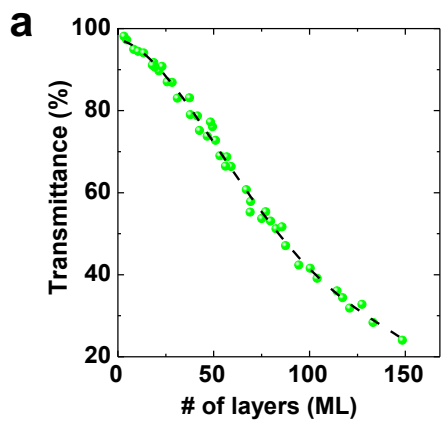 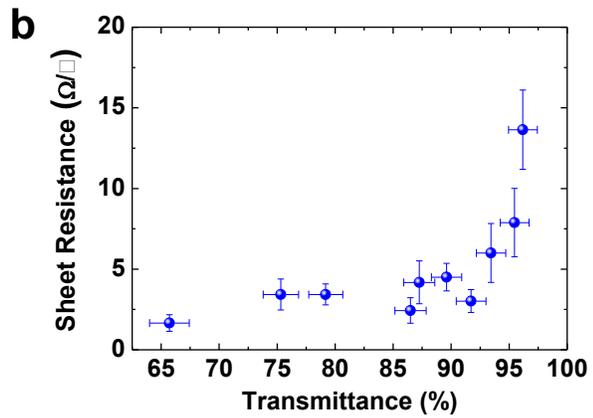

**Supplementary Figure 11**. For LiC$_6$ samples (a) polynomial fitting of transmittance *vs*. number of layers (b) correlation of sheet resistance vs. transmittance.

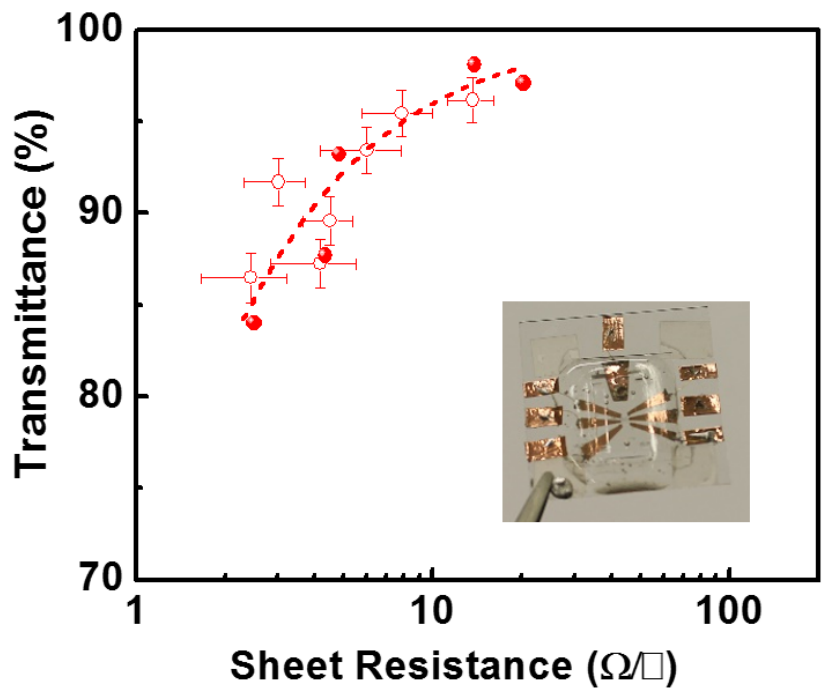

**Supplementary Figure 12.** Optical transmittance *vs*. sheet resistance for two batches of devices. Solid data points are taken on the same device using a concurrent optical/electrical measurement setup.

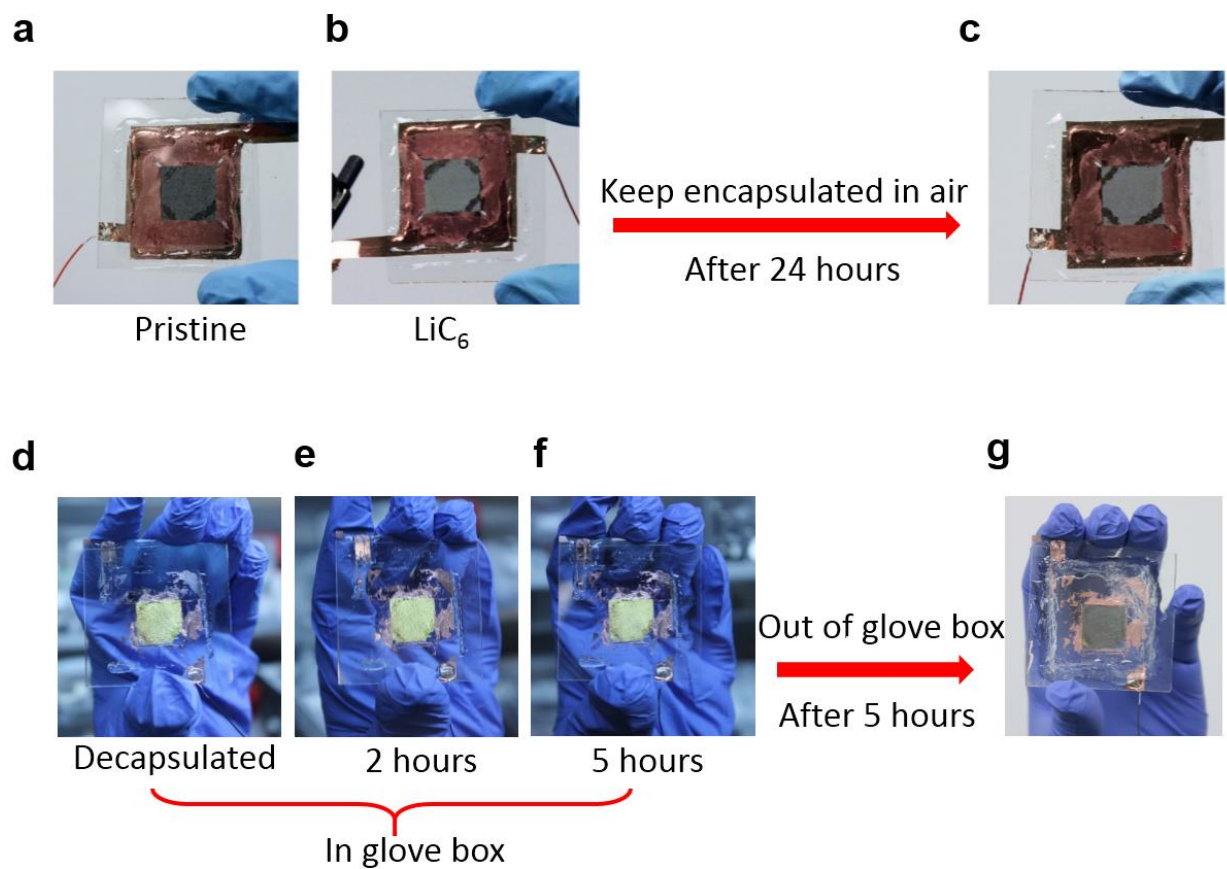

**Supplementary Figure 13**. Stability of encapsulated CVD thin graphite device in air. **d-g**, Decapsulated thick $LiC_6$ is stable in glove box but unstable in air.

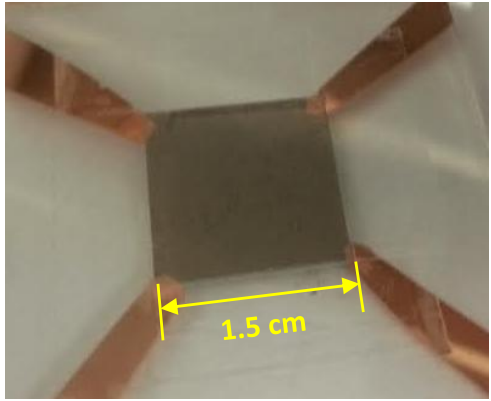

| | $R_s$ (Ω/□) (graphite) | $R_s$ (Ω/□) (LiC$_6$) |
|---|---|---|
| Dev 1 | 35.4 | 3.0 |
| Dev 2 | 47.7 | 3.9 |
| Dev 3 | 57.0 | 1.7 |

**Supplementary Figure 14**. Van de Pauw measurement of CVD ultrathin graphite and $R$s of pristine graphite and LiC$_6$ for three 40-nm-thick CVD graphite devices.

# Supplementary Discussion

## I. *In situ* optical transmission measurements

An electrochemical workstation (Biologic SP-150) is used to control charge/discharge of the Li-ultrathin graphite nanobattery and measure the time-dependent potential on intercalation (lithiation) and de-intercalation (delithiation), as shown in Supplementary Fig. 1. The charge current and discharge current are 0.5 µA and 1.5 µA, respectively. The charge time during the first charge cycle is much longer than the subsequent cycles, indicating that a passive layer (solid electrolyte interphase; SEI) forms primarily during the first charge cycle. The charging time decreases during the following cycles because the passive layer protects lithiated ultrathin graphite from further side reaction[1]. We adjust the applied charge current by testing different values of current during the first charge cycle and select an appropriate value (0.1-2.0 µA, which mostly depends on the total amount of exfoliated ultrathin graphite sheets attached to the metal contact) to achieve clear voltage potential plateaus during the Li-ultrathin graphite intercalation. The total charge time is approximately inversely proportional to the applied charge current.

At the end of each charge cycle we keep the voltage below 10 mV for enough time to ensure a complete formation of $LiC_6$ state, especially for thick sheets. This low voltage between ultrathin graphite and Li metal could result in an irreversible Li plating[2,3] to the edge of ultrathin graphite sheets at the end of charge cycle, as seen in Fig. 2g and 2i in the main text.

White light was also used as a light source for the transmission measurement of intercalated ultrathin graphite. During the intercalation process the transmittance has a more moderate change (Supplementary Figure 2a) instead of clear transmittance plateaus (Fig. 3a in main text) measured by 550 nm light source. For relatively thin sheets (0-30 layers) the transmittance difference between LiC12 and LiC6 is also reduced when white light is used (Supplementary Figure 2b) due to the special wavelength dependence of transmittance (Fig 4b,c in main text). Therefore, a light source with 550 nm wavelength is better for observing the variation of transmittance at different intercalation stages.

We also observed that the variation of optical transmittance is highly reversible by charging and discharging the Li-ultrathin graphite planar nano-battery. The transmittance (550 nm light source) switches between the values of pristine and LiC6 state as shown in Supplementary Figure 3. Here a relative large charge/discharge current of 1.0/2.0 µA is applied during the cycling to reduce the cycle period.

## II. *In situ* Raman measurements

A commercial micro Raman spectrometer (Labram Aramis model manfuctured by Horiba Jobin Yvon) is used for *in situ* measurements. The grating is 600 gr/mm, the laser source is a 633 nm He-Ne laser with 9mW power, and a D1 filter is used so the actual power is 0.9 mW.

The Fig. 2i shown in the main text can be used to confirm the stages of Li intercalation. The black line depicts the initial Raman spectra of an ultrathin graphite sample at a voltage of 1.00 V (vs. Li/Li$^+$), for which the Raman shift shows a typical G peak of graphene at 1580 cm$^{-1}$. The G peak shifts upward to 1600 cm$^{-1}$ at dilute Li intercalation stage (LiC$_{72}$), where the upshift due to Li doping[4]. When stage IV (LiC$_{36}$) starts to form, the G peak splits into two (1576.2 cm$^{-1}$ and 1601.8 cm$^{-1}$), which represents the interior and bounding layer modes of Li intercalated ultrathin graphite[5]. Upon further intercalation, the upper shifted peak grows while the lower peak vanishes, indicating a dominating bounding layer mode and the formation of the stage II intercalation compound (LiC$_{12}$). Finally, the two G peaks completely disappear, indicating the formation of Stage I (LiC$_6$), which can be simply understood as the Pauli blocking of the interband optical transition, hence there is no resonant Raman process[6]. The Raman spectra for Li-intercalated ultrathin graphite sheets agree well with previous studies of bulk samples[5], therefore, Raman microscopy can be used as one of the tools for differentiating lithiation stages of ultrathin graphite.

We also obtained Raman spectra over a wider range of Raman shift of 1200 - 3000 cm$^{-1}$ from a different sample, as shown in Supplementary Figure 4. The absence of D peak (1345 cm$^{-1}$) indicates that no degradation of the crystalline quality of ultrathin graphite occurred during electrochemical cycling. We also observed a vanishing of the 2D peak, and before the vanishing, the 2D frequency shifted from 2687.75 cm$^{-1}$ (pristine graphite) to 2677.73 cm$^{-1}$ (dilute stage) and then 2604 cm$^{-1}$ ($LiC_{36}$).

**III. Color change of ultrathin graphite during Li intercalation detected by reflective microscopy.**

The same band structure effects that give rise to wavelength-dependent transmittance also impact graphene reflectance, hence graphite changes color during Li intercalation process[7]. By combining the *in situ* optoelectronic measurement system with a reflective microscope (Olympus STM6), we also observed a color change of ultrathin graphite sheets during the Li-intercalation process. A series of digital pictures were captured with a color CCD camera (Olympus ColorView I). The incident white light source is LG-SP2 and intensity of light was kept constant during image capture. While charging the Li-ultrathin graphite nano-battery, ultrathin graphite sheets undergo a color change during the Li intercalation process. Three representative images are shown in Fig. S5, where the color can be seen to alter from white grey of the pristine state to dark blue of $LiC_{12}$ and finally golden yellow of $LiC_6$. Such color change could be partially explained by changes of optical transmission[8] (see Fig. 4 in main text), however, the color change is also highly dependent on the thickness of ultrathin graphite, as shown in Supplementary Figure 5. Therefore, the details of the reflectivity spectra as a function of Li concentration and layer thickness are worth exploring in a future study. Still, the lithiation interface (see Fig. 3c-g in main text) is not as clear as that detected by transmission microscopy.

**IV. Fabrication of planar nanobattery device for electro-optical measurement.**

Supplementary Figure 6 demonstrates the large, uniform thickness of ultrathin graphite flakes by mechanical exfoliation method. The large area uniformity ensured the reliability of the electrical transport measurement of our experiments.

The process flow chart in Supplementary Figure 7 illustrates the fabrication details of our multi-functional planar nanobattery device. The transfer method we used is similar to the one developed by Zomer et al.[9] Supplementary Figure 8 further shows large uniform ultrathin graphite transferred on Hall bar.

**V. Optical modeling details of Li intercalated ultrathin graphite**

The band theory for LiC$_6$ reveals a Dirac spectrum with a Fermi energy of ~ 1.5 eV.[10] The optical properties of Li intercalated ultrathin graphite can be simply modeled as doped ultrathin graphite with a 1.5 eV Fermi energy. There are two contributions to the optical conductivity or the dielectric function, a Drude free carrier term and an interband term, i.e., $\sigma(\omega) = \sigma_d + \sigma_{ib}$.

The Drude conductivity can be written as $\sigma_d = \dfrac{e^2 E_F N}{\pi \hbar (\gamma - i\omega)}$, where $\gamma = 1/\tau$ is the carrier relaxation rate and N is the number of layers. For $\hbar\omega > 2E_F$ the interband conductance turns on and its real part is given by $\operatorname{Re}\sigma_{ib} = \dfrac{\pi e^2 N}{2h}$; however, this step in $\sigma(\omega)$ produces a non-zero imaginary part given by $\operatorname{Im}\sigma_{ib} = -\dfrac{e^2 N}{2h} \ln\left|\dfrac{2E_F + \hbar\omega}{2E_F - \hbar\omega}\right|$. More accurately, thermal broadening of

the Pauli blocking leads to $\text{Re}\,\sigma_{ib} = \dfrac{\pi e^2 N}{2h}\left[\tanh\left(\dfrac{2E_F + \hbar\omega}{4kT}\right) + \tanh\left(\dfrac{2E_F - \hbar\omega}{4kT}\right)\right]$, and the imaginary part of $\sigma_{ib}$ is given by the Kramers-Kronig relations.

The complex dielectric function is also obtained from the optical conductivity by $\varepsilon(\omega) = \varepsilon_1 + i\varepsilon_2 = \varepsilon_\infty + i\dfrac{4\pi}{\omega}\sigma(\omega)$, where $\varepsilon_\infty$ is the high frequency dielectric constant ($\omega \approx W$, where $W$ is the bandwidth). The Pauli blocking edge produces a positive contribution to $\varepsilon_1(\omega)$, which, because of the negative free carrier contribution, leads to a plasma edge when $\varepsilon_1 = 0$ near the Pauli blocking edge. The real part of the conductivity from this model was shown schematically in Fig. 4g of the main text. The $\varepsilon_1 = 0$ also implies $\sigma_2 \approx 0$ near the maximum in the transmission, which allows the approximation of $\sigma_{opt}$ real in the Figure of Merit analysis. In Supplementary Figure 12 we show $\varepsilon_1(\omega)$ from the optical model. The $\varepsilon_1 \approx 0$ corresponds to the onset of transmission for bulk materials. Calculations of the transmission in our model conductivity and using the full slab transmission formulas accounts for the sharpening of the peak in the transmission curves for thicker films shown in Fig. 4c in the main text.

## VI. Electrical measurement at different Li-intercalation stages

The sheet resistance of ultrathin graphite at different intercalations stages is measured by a linear fit of four-probe IV curves; therefore, the contact resistance could be excluded. Typical curves are shown in Supplementary Figure 9.

During the Li intercalation it is difficult to measure the sheet resistance of pure $LiC_{12}$ state accurately because such measurement takes certain amount of time, while Li-intercalation is ongoing. Therefore, we measure the sheet resistance at the end of the transition from $LiC_{18}$ to $LiC_{12}$ so that $LiC_{12}$ is the majority component during the electrical measurement.

**VII. Correlation between sheet resistance and optical transmittance of LiC6**

It is difficult to measure optical transmittance and sheet resistance simultaneously for $LiC_6$ stage because most of the electrical measurements were performed on a probe station without optical characterization access. In order to correlate sheet resistance with transmittance, we first use polynomial fit (4 orders) to correlate transmittance with layer number (Supplementary Figure 8a): $y = a + bx + cx^2 + dx^3 + ex^4$ where $a = 97.43486$, $b = -0.10851$, $c = -0.01274$, $d = 1.10858 \times 10^{-4}$ and $e = -2.86171 \times 10^{-7}$. Thus, we can convert the layer number to transmittance of $LiC_6$ for the ultrathin graphite devices studied in electrical measurements, and the results are plotted as sheet resistance *vs.* corresponding transmittance in Supplementary Figure 10.

**VIII. Concurrent measurements of optical transmittance and sheet resistance**

By attaching an electrical probing setup to the transmission optical microscope, transmittance and four-probe resistance can be measured simultaneously on the same sample. Such data is plotted as solid circular points in Supplementary Figure 11, which locate very close to the original data, supporting the validity of the previous correlation.

**IX. Stability study of intercalated large area CVD ultrathin graphite**

In order to determine the stability of large scale $LiC_6$ films for transparent electrode application, we also carried out an initial stability test using commercially-obtained ultrathin graphite films grown by chemical vapor deposition (CVD), as shown in Supplementary Figure 13. After epoxy encapsulation, a CVD ultrathin graphite device was lithiated to $LiC_6$ state and stayed stable in ambient condition for over 48 hours, as shown in Supplementary Figure 13 a-c. We also confirmed that $LiC_6$ is unstable if exposed in air, as shown in Supplementary Figure 13 d-g. Therefore we conclude that once the device is appropriately sealed it is stable for transparent electrode applications. Again, standard industrial sealing techniques as used for Li-ion batteries should suffice to produce stable $LiC_6$ films. Supplementary Figure 14 shows the sheet resistance measurement using Van der Pauw method, and results of *Rs* for three 40-nm-thick CVD graphite devices.

## Supplementary References